\documentstyle[prl,aps,epsf,twocolumn]{revtex}

\tighten

\begin{document}
\draft
\twocolumn[\hsize\textwidth\columnwidth\hsize\csname @twocolumnfalse\endcsname

\title{\bf Numerical studies of the vibrational isocoordinate rule in
chalcogenide glasses}

\author{Normand Mousseau~\cite{mousadd} and D. A. Drabold\cite{drabadd}}               

\address{
Department of Physics and Astronomy, 
Condensed Matter and Surface Science Program,
Ohio University, Athens, OH 45701
}

\date{\today}

\maketitle

%\begin{center}
%\hspace{-11pt}
%{\bf Submitted to the Physical Review B}
%\end{center}

\begin{abstract}
Many properties of alloyed chalcogenide glasses can be closely
correlated with the average coordination of these compounds. This is the
case, for example, of the ultrasonic constants, dilatometric softening
temperature and the vibrational densities of states. What is striking,
however, is that these properties are nevertheless almost independent of
the composition at given average coordination. Here, we report on some
numerical verification of this experimental rule as applied to
vibrational density of states.  
\end{abstract}


\vskip2pc ]

\vspace*{-0.5cm} \narrowtext

\section{Introduction}

Establishing the microscopic properties of disordered materials based on
macroscopic probes is a difficult endeavor: the charateristic isotropy
of these materials limits measurements to mostly scalar,
orientation-averaged properties, reducing significantly the amount of
information accessible compared with, for example, what is available in
crystals. 

This is the case for scattering experiments. X-ray provides, after
Fourier transform, only an isotropic radial distribution function. This
smooth, structureless curve beyond medium-range order,  can be reproduced
numerically with a wide range of mutually inconsistent models. Such
measurement can at most provide a way of eliminating bad models, but is
useless as a tool for the positive identification among the other ones:
any model that fails to produce a realistic RDF is clearly incorrect,
this leaves, as shown using reverse Monte-Carlo, a wide range of
incompatible models.\cite{rmc}

The experimental evidence for an isocoordinate rule in chalcogenide
glasses provides yet another example of the difficulty of extracting
microscopic information from these disordered materials. This rule
states that for a given average coordination, samples with varying
compositions will display identical properties. The isocoordinate rule
has already been noted for a wealth of mechanical and thermal
properties such as ultrasonic elastic constants, hole relaxation, and
glass transition temperatures and hardness, and was found to hold for
the more complex vibrational density of
states~\cite{effey99,effey,effey99b}:
systems as different as Se$_{40}$As$_{60}$ and
Se$_{55}$As$_{30}$Ge$_{15}$, with an average coordination of
$\langle r \rangle =2.6$, show a similar VDOS in the transverse accoustic (TA)
region.

The vibrational isocoordinate rule (VIR) has, until now, only been
checked experimentally, with the inherent limitations due to atomic
species available and glass phase diagrams. This leaves a few questions
open regarding the range of validity of this rule as well as its
accuracy. In this paper, we present some results on a set of idealized
models that provide a bound on these two questions. 

Analytical study of this type of problem is difficult. Due to the
vectorial nature of the problem, even simplified topologies such as the
Bethe lattice are difficult to treat meaningfully. With the additional
topological disorder,  analytical solutions are beyond reach.
We report here on the results of direct numerical
simulations on a model system with simplified dynamics.

\section{Details of the simulation}

The simulations proceeded as follows. We start with a 4096-atom cell
of Sillium -- a perfectly tetravalent continuous-random network --
constructed by Djordjevic {\it et al.}\cite{djordjevic95} following
the prescription of Wooten, Winer and Weaire\cite{www}; this
network provides an idealized model with the appropriate initial
topology.  We then remove bonds at random in the network until we
reach the desired concentration of 2-, 3- and 4-fold atoms.  In the
first stage of the simulation, we do not enforce any extra correlation
and the final network corresponds to a perfectly random amorphous
alloy. We then relax the network using a Kirkwood potential with
interactions based on the table of neighbours, not on distance.
\begin{equation}
E = \frac{\alpha}{2} \sum_{\langle i j \rangle} (L_{ij} - L_{0})^2 +
\frac{\beta}{8} L_0^2 \sum_{\langle ijk \rangle} \left( \cos
\theta_{jik} + \frac{1}{3} \right)^2
\end{equation}
where $\alpha$ and $\beta$ are taken to be the same for {\it all} bonds
and $L_0$ is the ideal bond length.
We take a ratio of three-body to two-body force, $\beta/\alpha=0.2$,
typical of tetrahedral semiconductors \cite{cai92}. 

The resulting network is one of identical atoms except for the
coordination. This is not too far from SeAsGe
chalcogenides; because they sit side by side in the same row, these
elements share very similar masses, elastic properties and Pauling
electronegativities. As an
additional simplification, we take the same tetrahedral angle for all
triplets in the network. Real Se and As, in the respective 2- and 3-fold
configuration, have angles that deviate from this value and tend towards
120 degrees. This simplification is less drastic than it appears because
of the relatively low coordination, allowing a significant degree of
flexibility in the network: angles can then be accomodated at very
little elastic cost.  A more serious concern is that although the 3
elements have very similar elastic constants in 4-fold environment, then
bonding will get stronger as the coordination decreases. Comparison
with experimental data shows that this effect shows up mostly in the
high frequency TO peak. Moreover, because of the square root scaling,
the deviation becomes apparent only between samples at the extreme of
the composition scale. 

To verify the isocoordinate rule, we prepare 3 different compositions at
each average coordination from $\langle r \rangle =2.0$ to $3.0$ (except
at $\langle r \rangle =2.0$,
where only two different cells are created). We then proceed to
distribute at random a desired proportion of 2-, 3- and 4--fold atoms. The
three configurations typically correspond to (1) a configuration with a
maximum of 3-fold atoms for the given average coordination, (2) one with
a maximum of 4--fold, and (3) a composition between the two.  For
example, at $\langle r \rangle =3.0$, we create a configuration with 50 percent of
2--old and 50 percent of 4--fold atoms, one with 25, 50 and 25 percent
of 2-, 3-, and 4--fold atoms, respectively, and one with 100 percent of
3-fold atoms.  This gives us the widest spectrum possible 
to study the VIR.   Because we are not constrained by the glass forming
diagram, this is also  wider than what can be achieved experimentally. 

After the topology has been determined, each sample is relaxed with
the Kirkwood potential, using periodic boundary conditions.  The
dynamical matrix is then computed numerically on the fully relaxed
configuration.  This 12288 $\times$ 12288 matrix is then diagonalized
exactly in order to obtain the full vibrational properties. The
eigenvalues are binned and smoothed with a Gaussian of experimental
width to provide the vibrational density of states presented in this
paper.

We also introduce some chemical correlations to see how sensitive the
VIR is to local fluctuations. We study here two types of correlations:
phase separation -- intoducing some king of homopolar preference -- and
mixing, with heteropolar bonds. A cost function is introduced in the
bond-distributing sub-routine and all other phases of simulation remain
the same.  

\section{Results and discussion}

\subsection{The vibrational isocoordinate rule}

\begin{figure}

\epsfxsize=16cm
\epsfbox{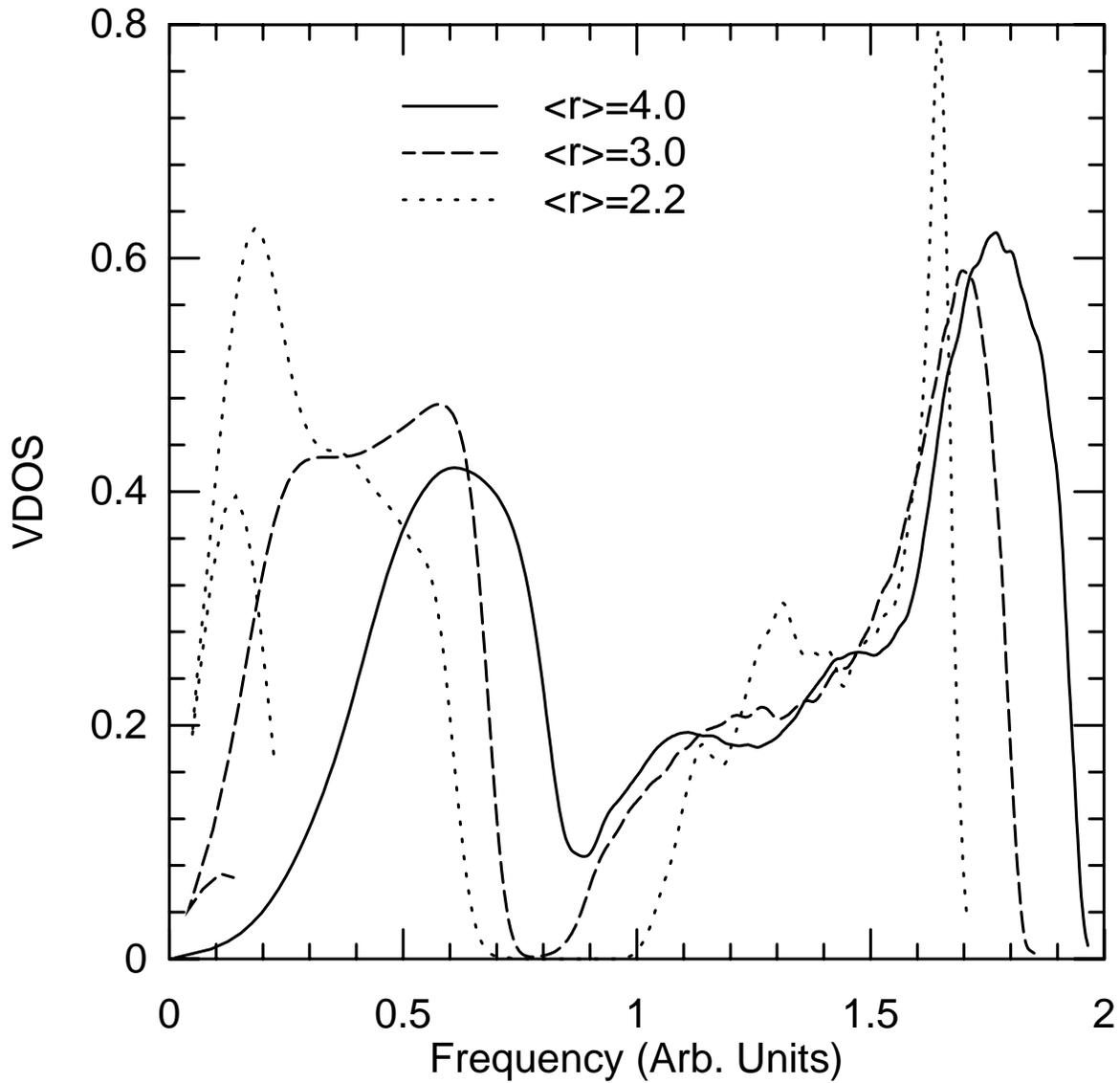}

\vspace{2cm}

\caption{Vibrational density of state (VDOS) as a function of frequency for
a set of configurations of different composition and average
coordination. Each plot shows the VDOS for two or three different
composition but identical average coordination. The compositions are
given in table \ref{table1}. In each plot, the choice of line goes as
a function of increasing percentage of 2-fold atoms: solid, dashed and 
dotted. The additional peak at low energies in this and later figures is
due to floppy modes.}
\label{fig:vdos}
\end{figure}

Figure \ref{fig:vdos} shows the vibrational density of state as a
function of average coordination  from $\langle r \rangle = 2.0$ to
$3.0$. This distribution goes through the topological rigidity
transition at $\langle r \rangle = 2.4$. First, we note that the VIR is
approximately valid for two frequency bands: the transverse accoustic
band -- below $f=0.7$ and transverse optic band -- above $f=1.5$.  This
holds for configurations with significant difference of composition,
even configurations as different as the 100 percent 3-fold vs. 50-50 of
2- and 4--fold, show fairly similar VDOS in these regions. This is a
wider application range than what was measured experimentally; the data
reported by Effey and Cappelletti  shows good overlap for the TA band
but no consistent overlap in the higher frequency region of the
VDOS. This is especially true of samples with widely different
composition. The experimental shift in the TO peak seems to follow the 
concentration of Se. This is consistent with the expected increase in
stiffness of the low coordinated atoms discussed above.

Taking this into account, a second look at the figures indicates
clearly that the VIR is not an exact law.  The structure of
the TA peak in the configurations with $\langle r \rangle = 3.0$ shows
significant variations following the concentration of 3-fold
coordinated atoms. This shoulder, slightly above $f=0.5$, decreases in
importance with the average coordination.  This effect is not seen
experimentally; the range of compositions for the real samples,
however, is much narrower than that studied here.

Figure \ref{fig:vdos} also provides some indication about the relation
between specific structures in the VDOS and the local
environment. Because the VIR holds well, these are features that
depend mostly weakly on the details of the composition and cannot be
of much help in experimental situations. We have already mentioned the
presence of additional modes related to 3-fold coordinated atoms on
the high-frequency side of the TA band -- a feature that is also
present at high coordination in experimental measurement. Although
this shoulder is significant, it represents the maximum impact it can
have, with a direct comparison between a fully three-fold coordinated
sample and one with zero such atoms. Given that the two network have a
totally different topology, it is the similarity between the two
curves rather than their differences that has to be emphasized.

\begin{figure}

\epsfxsize=20cm
\epsfbox{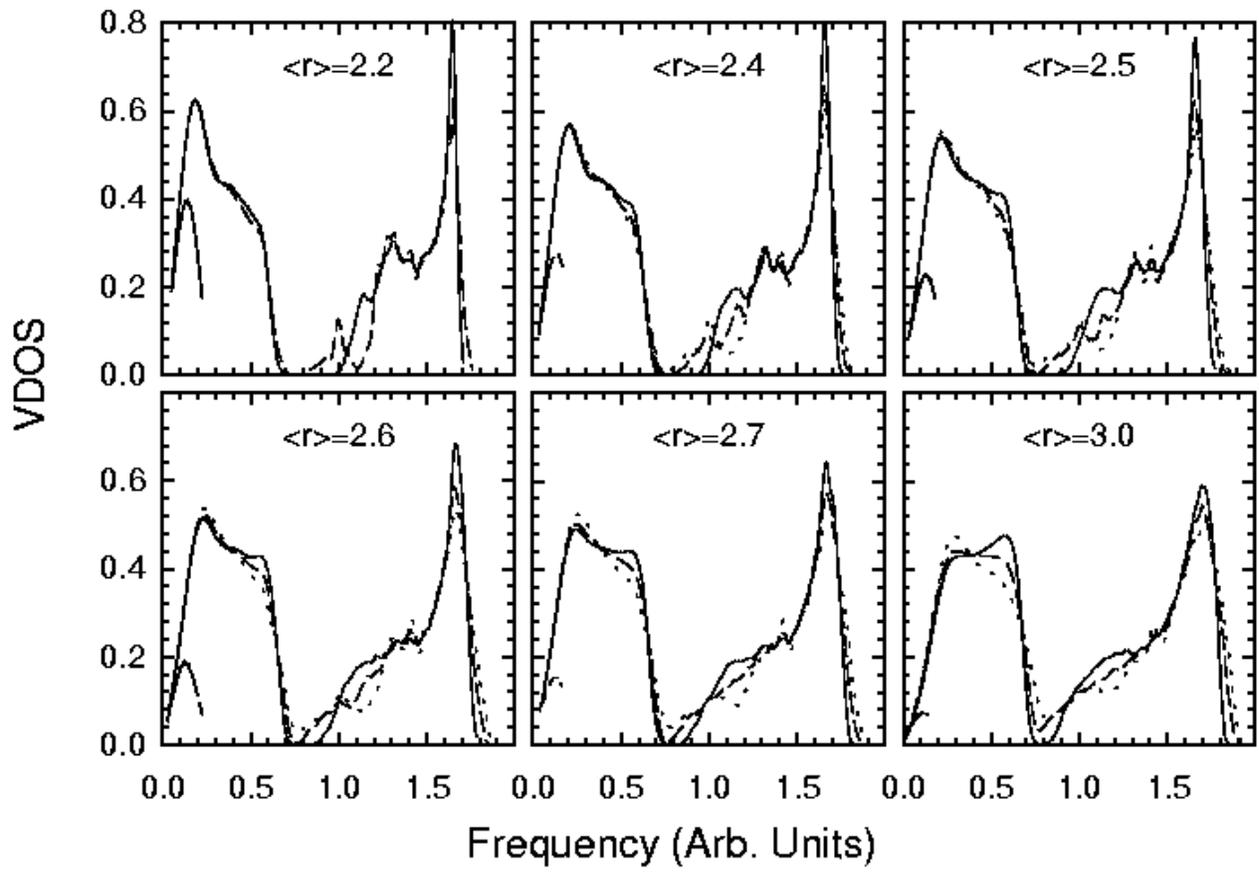}

\caption{Evolution of the VDOS as a function of average
coordination. The shift in the overal VDOS is explained by the
significant decrease in the rigidity of the network as the number of
bond diminishes.}
\label{fig:comp}
\end{figure}

The structure between the TA and the TO bands is more directly
sensitive to the details of local structure, especially at low average
coordination where more modes become localized. The structure around
$f=1$ consistently represents the four-fold coordinated atoms.  We can
compare these structures to a fully four-fold structure
(Fig. \ref{fig:comp}).

As we decrease the average coordination of the networks, we go through
the topological rigidity threshold, at $\langle r \rangle =
2.4$. Below this value, the network becomes floppy and its macroscopic
elastic constants vanish; local rigidity remains, however, and the
VDOS is mostly unaffected except for a shift in the position of the
peaks and an accumulation of modes at low frequencies. The signature
of these zero-frequency modes is reported here in the backward peak
formed at low frequencies and the accumulation at the lower-end of the
TO peak in networks with large fraction of 2-fold coordinated
atoms. The backward peak corresponds to spurious imaginary frequencies
associated with floppy modes.  Based on the theory of topological
rigididity, these modes are localized above $p_c$ and span the whole
network below this threshold.

The floppiness of the network is also reflected in the TO peak.
Strikingly, the width of this peak is much more related to the
overall coordination of the network than with local environment. From
the work of Alben and Weaire, the TA peak has been associated with the
overall coordination while the TO peak had been ascribed to the local
tetrahedral symmetry. Vibrational density of states of fully four-fold
cells generally gives a much wider if somewhat lower TO peak with, in
the case of ill-coordinated networks, a very flat
structure.\cite{mousseau91} Such a wide peak has generally been
associated with non-tetrahedral environment but it is clear, based on
the results obtained here, that the present of strain is also
necessary. Even at $\langle r \rangle =3.0$ the network contains
enough floppy modes to relax a good part of the strain: the optical
modes become then more localized and emerge as a relatively sharp
feature at the edge of the vibrational spectrum.

As the averge coordination decreases the LA (around $f=1.0$) and LO
($f=1.4$) features also become more prominent. The first is
particularly sensitive to the concentration of 4-fold atoms. This is
especially clear at the lowest average coordination where this feature
is totally absent in the Se$_{80}$As$_{20}$ sample. At the same time,
the LA peak is shifted upward to about $f=1.15$ and can therefore be
associated with the presence of 3-fold atoms.  This interpretation is
in full agreement with the experimental results reported by Effey and
Cappelletti\cite{effey99}.

\subsection{Correlations}

\begin{figure}

\epsfxsize=16cm
\epsfbox{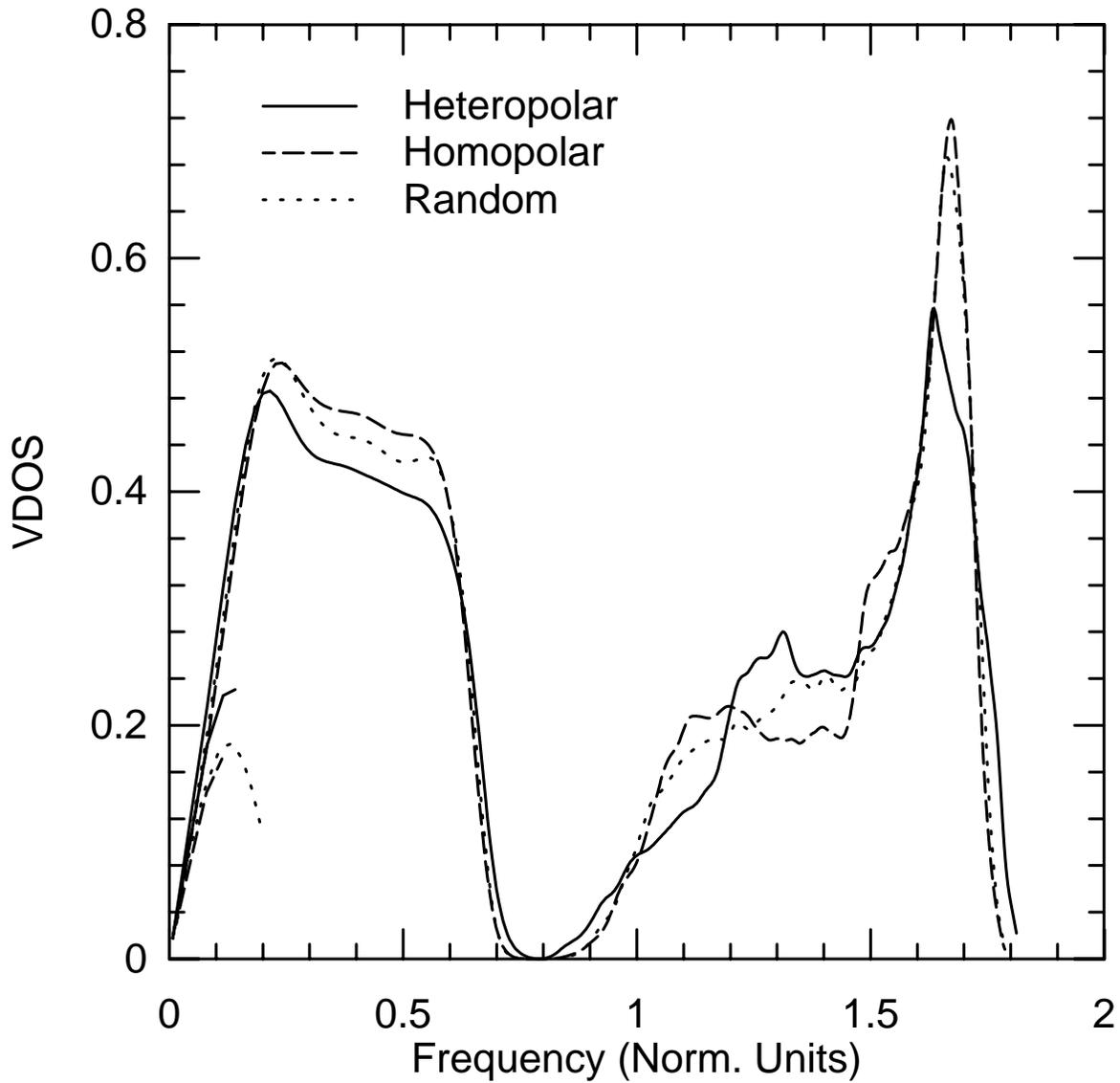}

\vspace{2cm}

\caption{Effect of short-range correlations on the vibrational density 
of state. We show here three samples at $\langle r \rangle = 2.6$ with 
40 percent 2-fold and 60 percent 3-fold atoms but different local
chemical order. Bonds between similar atoms are highly favored in the
homopolar sample and highly penalized in the heteropolar sample. No
account of chemical cost is introduced in the random case.}
\label{fig:correl}
\end{figure}

The above results are all for non-correlated samples. It is not
impossible, however, that chemical ordering might take place in
chalcogenide glasses. To address this question, we prepared, at
$\langle r \rangle =2.6$, two configurations of nominal
Se$_{40}$As$_{60}$ with full mixing -heteropolar bonds fully favored
-- and demixing -- homopolar bonds preferred. In the first case, no
2-fold atom is bonded to another 2-fold atom. In the second case,
clustering tends to take place. Figure
\ref{fig:correl} shows the VDOS for these two samples as compared with
the non-correlated case already described in the previous section. 

There is remarkably little impact on the spectrum. Although we see
some broadening of the TA peak and a sharp structure at the far left of
the TA peak in the demixed case, due to the very floppy selenium chains,
the rest of the spectrum is essentially unchanged. Except, that is, for
some shifting in the two peaks associated with the LA and LO
peaks. Mixed environment seem to contribute more to the peak at 1.15
while demixed environments shift the peak to higher frequencies. The
average of these two effects is manifest in the non-correlated
spectrum.  These changes in the spectrum are minor, however, and too
subtle to allow any quantitative extraction from experimental data. 

The choice of $\langle r \rangle=2.6$ was based on some peculiar
spectra found experimentally for Se$_{2}$As$_{3}$.\cite{effey99}
Although we have been unable to reproduce the experimental behavior,
we can conclude from this section that simple two-body correlation is
not sufficient to provide the type of structure in the VDOS seen
experimentally. More striking local changes need to occur, such as
pseudo-molecular constructions, which give rise to localized
modes\cite{effey99}.

\section{Conclusions}

Following some interesting experiments displaying an isocoordinate rule
for chalcogenide glasses, we studied the case of an ideal ternary
glass consisting of identical atoms save for their coordination: same
mass, same elastic constants same angular forces. It is then possible to
examine the nature of this rule without interference from any other
contribution to the vibrational density of state. 

We have examined the isocoordinate rule at average concentrations
varying from 2.0 to 3.0. What we find is that the isoccordinate rule is
not an exact one. However there is much less variation of the
vibrational density of state for a given average concentration, even as
the type of atom is varied completely, with, for example, from totally
three-fold to a mixture containing only 2 and 4-fold atoms than changing
the overall concentration of atoms by 0.1 or 0.2.

Moreover, it appears that the VDOS is not very sensitive either to the
presence of correlations. Changing considerably the local correlations
affects very little even the qualitative structure of the VDOS.

All these results, in parallel with the experimental results obtained
recently point to the fact that it is difficult to obtain
positive information simply from the VDOS: if two samples have a
different VDOS, it is clear that they diverge; nothing can be said,
however if they have the same VDOS.

\section{Acknowledgements}

We thank Drs. Ronald Cappelletti and Birgit Effey for many stimulating
discussions.  NM thanks the NSF for support under Grant DMR 9805848,
DAD thanks NSF for support under grants DMR 9618789 and DMR 9604921.

\bibliographystyle{prsty}

\begin{table}
\caption{Compositions used in this paper. All samples are created by
removing bonds from a perfectly coordinated 4096-atom configuration
while enforcing a certain proportion of 2-, 3- and 4-fold atoms.}
\label{table1}
\begin{tabular}{cccc}
Average composition & 2-fold & 3-fold & 4-fold \\ \hline 
2.2   & 0.80 & 0.20 & 0.00 \\
      & 0.90 & 0.00 & 0.10 \\ 

2.4   & 0.60 & 0.40 & 0.00 \\
      & 0.70 & 0.20 & 0.10 \\
      & 0.80 & 0.00 & 0.20 \\ 

2.5   & 0.50 & 0.50 & 0.00 \\
      & 0.65 & 0.20 & 0.15 \\
      & 0.75 & 0.00 & 0.25 \\ 

2.6   & 0.40 & 0.60 & 0.00 \\
      & 0.55 & 0.30 & 0.15 \\
      & 0.70 & 0.00 & 0.30 \\ 

2.7   & 0.30 & 0.70 & 0.00 \\
      & 0.45 & 0.40 & 0.15 \\ 
      & 0.65 & 0.00 & 0.35 \\ 

3.0   & 0.00 & 1.00 & 0.00 \\
      & 0.25 & 0.50 & 0.25 \\ 
      & 0.50 & 0.00 & 0.50 
\end{tabular}
\end{table}

\twocolumn[\hsize\textwidth\columnwidth\hsize\csname @twocolumnfalse\endcsname

\vskip2pc ] 
\newpage
\twocolumn[\hsize\textwidth\columnwidth\hsize\csname @twocolumnfalse\endcsname

\vskip2pc ] 
\newpage
\twocolumn[\hsize\textwidth\columnwidth\hsize\csname @twocolumnfalse\endcsname

\vskip2pc ] 

\end{document}